\documentstyle[preprint,tighten,epsf,eqsecnum,aps,floats]{revtex}

\begin{document}
\def\be{\begin{equation}}
 \def \ee{\end{equation}}
\def\bea{\begin{eqnarray}}\def\eea{\end{eqnarray}}
\def\eqn {Eq.~(\ref }

\preprint{NT@UW-01-14} 
\title{
  Return of the  EMC  Effect}
\author{Gerald A. Miller and Jason R. Smith}
\address 
{Department of Physics\\
  University of Washington\\Seattle, WA 98195-1560}
\maketitle
\begin{abstract}
The relationship between the properties of nuclear matter and
structure functions measured in lepton-nucleus deep inelastic
scattering is investigated using light front dynamics. We find
that relativistic mean field models such as
  the Wale{c}ka, Zimanyi-Moszkowski (and point-coupling versions of the same)
  and Rusnak-Furnstahl  models contain essentially
  no binding effect, in accord with an earlier calculation by Birse.
These models are found to obey
   the Hugenholtz-van
Hove theorem, which is applicable if nucleons are the only degrees of freedom.
Any model in which the entire Fock space wave function can be represented in
  terms of free nucleons must obey this theorem, which implies that all of the
  plus momentum is carried by nucleons, and therefore that there will be
  essentially no binding effect. The explicit presence of nuclear mesons
  allows one to obtain a modified form of the Hugenholtz-van Hove theorem,
  which
  is equivalent to the often-used momentum sum rule.
  These results argue in favor of a conclusion that
  the depletion of the deep inelastic structure function
  observed in the valence quark regime is due to some interesting effect
  involving dynamics beyond the conventional nucleon-meson treatment of nuclear
  physics.

miller/jasons/reorder.tex
\end{abstract}

\maketitle
\section {Introduction}
The European Muon Collaboration (EMC)
effect in which the structure function of a nucleus, measured in
deep inelastic scattering at values of Bjorken $x\ge 0.4$
corresponding to the valence quark regime,
was found to be   reduced  compared with that of a free nucleon was discovered
almost twenty years ago\cite{EMCrefs}. Despite much experimental and
theoretical
progress\cite{EMCrevs,fs2},
no unique and universally accepted explanation of the
depletion has emerged. The immediate
parton model interpretation  that the nucleon
bound in a nucleus carries less
momentum than in free space seems uncontested, but
determining the
underlying origin remains an elusive goal.

One popular explanation is that
conventional nuclear binding effects are responsible. The conventional
lore is that
 the nuclear structure function $F_{2A}(x)$ (which gives the momentum
distribution of a quark in a nucleus as a function of the fractional momentum
carried)
can be obtained from the light front distribution function
$f(y)$  (which gives
 the probability that a nucleon
carries a fractional
momentum $y$) and the nucleon structure function of a free nucleon
$F_{2N}$ using  the relation\cite{simple}:
\begin{equation}
{F_{2A}^{\rm lore}(x)\over A}=\int dy f(y) F_{2N}(x/y). \label{deep}
\end{equation}
This formula has a simple interpretation as an expression which gives
a probability as a sum of products of probabilities. The variable
$x$ is the Bjorken variable
$x=Q^2/2 M \nu$, and 
 $y$ is the $A$ times the fraction of the
nuclear  plus-momentum carried by the nucleon, The plus component
of a four vector is the sum of the time and third spatial
component, so if $k^\mu$ is the momentum of a nucleon and $P^\mu$
is the momentum of the target nucleus
$y=(k^0+k^3)A/P^+=(k^0+k^3)A/M_A=(k^0+k^3)/\overline{M}$, in which
the nucleus is taken to be at rest with $P^+=M_A$. One can easily
use conventional nuclear physics to obtain the probability that a
nucleon carries a three momentum ${\bf k}$, but, if  one uses only
naive considerations, one faces a puzzle when deciding how to
choose the value of $k^0$. Should one use the average separation
energy, or the average nucleon mass $\overline{M}$, or possibly
the effective mass in the chosen many-body theory?

The essence of the  binding
explanation is that $k^0$ is given by the free nucleon mass
$M$ minus the average  separation energy $\epsilon$. Then $f(y)$
is  narrowly peaked at $y=1-\epsilon/M$
($\epsilon$ (with $\sim$
70 MeV for  infinite nuclear matter
\cite{Muther:2000qx}).
In this case,   the structure
function of a bound  nucleon is approximately obtained by
 replacing  $F_{2N}(x)$ by
$F_{2N}(x/(1-\epsilon/M))$. The increase in the argument leads to a significant
reduction in the value of the nuclear structure function
The theoretical understanding of the  binding effect (as of 1996)
is reviewed nicely in the book~\cite{Boffi}, which summarizes the various
treatments as ``not completely satisfactory''.
This kind of explanation seems
very natural because nuclear binding is known to
occur, so such an effect must be understood thoroughly before hoping to extract
information about a possible host of more interesting exotic effects.
In any case,  one needs to supply a derivation to avoid the need to
arbitrarily choose a prescription for $k^0$.

This need  drove one of us on to the light front
\cite{Miller:1997xh,Miller:1997cr}. That is,  to attempt to use light front
dynamics to derive the nuclear wave function. The reason for  this is
that, in the parton model
 $x$ is the ratio of the plus component of the momentum of the struck
quark to that of the target,   and it is the  plus component of
the momentum which was observed to be depleted by the EMC. In the
view of Ref.\cite{Miller:2000kv}, using light front dynamics is
the most effective way to assess the influence of binding effects.
However, one must pay the price of computing nuclear wave
functions using these dynamics.

The first attempts\cite{Miller:1997xh,Miller:1997cr} in this
direction employed the popular and successful Walecka
model\cite{Walecka:1974qa} which has many effective
descendants\cite{Serot:1997xg,Rusnak:1997dj,Furnstahl:2000rm}. The
salient result was that vector mesons carried 35\% of the nuclear
plus momentum and nucleons only 65\% ($P_N^+/P^+=0.65$), far
smaller than the value $(1-\epsilon/M)\sim0.95$ needed to
reproduce the observations for the iron nucleus. However, the
connection between the nucleon momentum distribution computed
using light front dynamics and that used in computing the deep
inelastic structure function was not made. Recently, the authors
of Ref.~\cite{HB} have claimed that quark distribution functions
are not parton probabilities. Their message to us is   that, in
any situation, one needs to derive the connection between the
constituent distribution function and the observed data. That work
stimulated us to undertake the present investigation in which we
derive the connection between the nucleon momentum distribution
and the structure function measured in deep inelastic scattering.

 First we outline our procedure.
We start in Section~\ref{ratiofunction} by  considering
relativistic  models of infinite nuclear matter computed using the
mean field approximation. We derive and apply  the nucleon
distribution function $f_N(y)$ appropriate for use in  computing
deep inelastic scattering structure functions. The  function
$f_N(y)$ is shown to be the one which  maintains the covariance of
the formalism, and in which the nucleons carry the entire
plus-momentum, $P^+$ of the  nucleus \cite{notation}. This result
is obtained independently of the specific relativistic mean field
theory used, so no such theory
contains
the binding effect discussed above. The only binding effect arises from
the average binding energy of the nucleus (16 MeV for infinite nuclear matter),
and is far too small to explain the observed depletion of the structure
function.
This is  in
accord with an earlier  similar
finding by Birse\cite{birse}. The generality of
this result encourages us to seek a broader context. This is found in
the  Hugenholtz-van Hove theorem\cite{HvH} which states that
the binding energy of the
level at the Fermi surface is equal to the average binding energy, or
the  energy of the level at the Fermi surface $E_F$ is equal
to the nuclear mass divided by A:
\be E_F=M_A/A\equiv \overline{M}.\label{hvh}\ee
 This theorem is the consequence  of using the
condition that the total pressure of the nucleus vanishes at equilibrium, and
the
assumption that nucleons are the only degrees of freedom contributing to
the nuclear energy. Thus this theorem is a signal that $P^+=P_N^+$ or that
nucleons account for the entire plus momentum of the nucleus. This generally
is understood to imply  that there will be no EMC binding effect\cite{fs2},
thus any model which obeys \eqn{hvh}) can be expected not to have one.

The next  step is to recall in
Section~\ref{walecka} how light front
dynamics is applied to computing the properties of infinite nuclear matter
using
the Walecka model (as a specific example) in mean field approximation (MFA).
The purpose is to
illustrate
the general formalism needed
to go beyond the mean field approximation,
provide an explicit
example of the general results presented in Section~II,  study the nuclear
structure origins
of the nuclear momentum content, and
  show explicitly that the Hugenholtz-van Hove theorem is satisfied.

 In
Section~\ref{four} we introduce four other
model Lagrangians in which the values of the effective mass and vector meson
field
vary widely. Again
our specific calculations  are limited to
the MFA. However, in Section~\ref{beyond},
the application of the Hugenholtz-van Hove theorem\cite{HvH}
allows us to make some general statements 
about models which include nucleon-nucleon correlations. In particular,
we use this theorem to explain  why no binding effect is contained
in
any model, such as that of Ref.~\cite{Pieper:2001mp},
in which nucleons are responsible for the entire plus momentum of the
nucleus. This is in accord with early observations of Ref.~\cite{fs2}, but now
there is an additional ability to compute all of the relevant nuclear
properties using light front dynamics. We also use our findings  assess
existing treatments of the binding effect.
Section~\ref{summary} is a  summary of   our  results and their
implications. 
In  Appendix~\ref{4app} we use light front dynamics to
compute nuclear properties of the  four models of
Section~\ref{four}.
\section{ Deep Inelastic Scattering From Nuclei}
\label{ratiofunction}

We are testing the hypothesis that conventional nuclear dynamics
can explain the EMC effect. This means that we need to include
possible binding energy and Fermi motion effects, but not
dynamics related to true modifications of the nucleon structure or
off-shell effects caused by the nuclear medium. The key
assumption is that the system formed by the absorption of the
photon 
is not a bound nucleon and therefore does not have the same interaction.
The relevant lifetime of the struck system is  ${1\over x M}\le 0.5$ fm
(for $x\ge0.5 $) which
 corresponds to a very short nuclear time, too short
for  interactions.
\begin{figure}
\centering
\epsffile{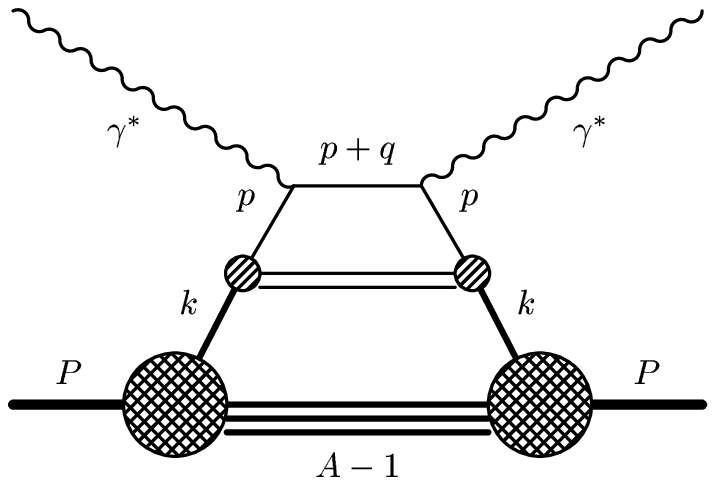}
\caption{Diagram for computing the nuclear structure function. A nucleus of
  momentum $P$ emits a nucleon of momentum $k$, which emits a quark of momentum
  $p$, which absorbs the virtual photon of momentum $q$.
  }
\label{fig:disdiag}
\end{figure}
In this case the use of  a manifestly
covariant formulation to derive the expression for the structure function
leads to a convolution formula.
If one uses the
free nucleon structure function (neglecting off-shell effects)
and  $F_{2N}$ for
free nucleons one finds\cite{jm}
  \bea
{F_{2A}(x_A)\over A}=\int^\infty_{x_A} dy f_N(y) F_{2N}(x_A/y), \label{deeper}\\
 f_N(y)=\int {d^4k\over (2\pi)^4} \delta(y-{k^0+k^3\over \overline{M}})
Tr\left[ {\gamma^+\over 2P^+A}\chi(k,P)\right],\label{good}\eea
where $P$ is the total four momentum of the nucleus, and
\be x_A\equiv Q^2A/2P\cdot q=x AM/ {M}_A=xM/\overline{M}\label{xa}\ee
with $M$ as the free nucleon
mass and $\overline{M}\equiv M_A/A$.
The function $\chi(k,P)$ is (proportional to)
 the connected part of
the nuclear expectation value of   the
nucleon  Green's function \cite{explain}, and the trace
is over the Dirac and isospin indices.
That $f_N(y)$ is a Lorentz scalar is manifest from the structure of
Eq.~(\ref{good}).
We note the appearance of $\overline{M}$ instead of the free value of the
nucleon mass $M$. This arises here from the definition (\ref{xa}) and
the feature that, in the Bjorken limit,
the nuclear structure function
depends on the ratio $p^+/P^+=(p^+/k^+)(k^+/P^+)$, where
$p^\mu$ is the quark momentum, with $P^+=M_A=A\overline{M}$.
The basis of the formula (\ref{good}) is that both the
quark and nucleon distributions are directly related to manifestly
covariant  Green's functions\cite{jm,jaffe}. This is a standard result using
nothing more than the stated assumptions and the Feynman diagram in  Fig.~1.

The manifestly covariant form of the single nucleon Green's function
has been known for a long time\cite{Serot:1986ey}, and
 its use (in the nucleus rest frame) leads to the result
\bea \chi(k,P)&=&-i2P^+\Omega\left(\gamma\cdot(k-g_vV)+M^* \right)\nonumber\\
&\times&\left[{1\over(k-V)^2-{M^*}^2+i\epsilon}
+{i\pi \over E^*(k)}\delta(k^0-E^*(k)-g_vV^0))\theta(k_F-\vert{\bf k}\vert)\right],
\label{gf}\eea
where
\be E^*(k)\equiv\sqrt{{M^*}^2+{\bf k}^2}.\ee
The general form of the Green's function depends on a vector potential
$V=(V^0,{\bf 0})$ for a nucleus at rest, and the effective mass
$M^*$ which includes
the effects of interactions on the nucleon mass. The values of $V$ and $M^*$
depend on the specific Lagrangian employed, but the form of the Green's
function is general. Recall also that ${V}^-=V^+=V^0$ for
 the expectation values of vector meson fields in the nucleus rest frame.

The result (\ref{gf})
 was first obtained using the conventional equal time approach,
but the very same can also be obtained from the light front formalism. In that
case it is
 necessary to include the effects of
the instantaneous part of the nucleon Green's function and those of
the instantaneous meson exchange.

The next step is to insert the connected part (second term) of  (\ref{gf})
into Eq.~(\ref{good}) for $f_N(y)$. This gives, after taking the
trace and using the delta function to integrate   over $k^0$, the result
\bea
f_N(y)={4\over (2\pi)^3\rho_B}\int d^2k_\perp\;dk^3 \frac{E^*(k) +k^3}{E^*(k)}
\delta(y-{E^*(k)+g_vV^++k^3\over
  \overline{M}})\theta(k_F-\vert{\bf k}\vert)
.\label{mid}\eea
The integration is simplified by using the transformation 
\bea
k^+\equiv E^*(k)+k^3,\label{k3}
\eea
which makes a connection with light front variables\cite{also}.
It is an exercise in geometry to show that the Fermi volume can be
re-expressed in terms of $k^+$ using
\be k_\perp^2+(k^+-E_F^*)^2\le k_F^2,\quad E_F^*\equiv
\sqrt{k_F^2+{M^*}^2},\label{fermisphere} \ee
so that Eq.~(\ref{mid}) becomes
\bea
f_N(y)={4\over (2\pi)^3\rho_B}\int d^2k_\perp \int dk^+
\theta\left(k_F^2-k_\perp^2-(k^+-E_F^*)^2\right)
\delta(y-{k^+ +g_vV^+\over\overline{M}})
.\label{preshift}
\eea
The use of   the definition of the
energy of a nucleon at the Fermi surface,
\be E_F= E^*_F+g_vV^+=E^*_F+g_vV^0,\ee
allows one to achieve a simple expression
for $f_N(y)$:
\be
f_N(y)= 
{3\over 4} {\overline{M}^3\over k_F^3}\theta((E_F+k_F)/\overline{M}-y)
\theta(y-(E_F-k_F)/\overline{M}))\left[
{k_F^2\over \overline{M}^2}-({E_F\over\overline{M}}-y)^2\right].\label{shiftm1}
\ee
The result \eqn{shiftm1})
can be further simplified by using the Hugenholtz-van Hove theorem
displayed in \eqn{hvh}).
Section~\ref{walecka} contains an explicit demonstration
 of \eqn{hvh}) for the Walecka model and the Appendix contains  a similar
 demonstration for the
 other relativistic models evaluated using  the mean field
approximation. Using \eqn{hvh}) in \eqn{shiftm1}) therefore
leads to the general result
\be
f_N(y)=
{3\over 4} {\overline{M}^3\over k_F^3}\theta(1+k_F/\overline{M}-y)
\theta(y-(1-k_F/\overline{M}))\left[
{k_F^2\over \overline{M}^2}-(1-y)^2\right],\label{shift}
\ee  correct  for  any
relativistic mean field theory of infinite nuclear matter. Different
theories with the same binding energy and Fermi momentum
may have very different  scalar  and vector potentials, but
must have the same $f_N(y)$.

A result very similar to Eq.~(\ref{shift})
 was previously obtained by Birse
\cite{birse}. The difference between his formula and ours is the appearance of
$\overline{M}$ in the function $f_N(y)$, whereas he uses $M$. This difference is
a small effect numerically, and therefore our  conclusions will be the
same
as his.

The baryon sum rule and momentum sum rules are derived by taking the first  two
moments of $f_N(y)$. This gives:
\begin{eqnarray}
\int dy f_N(y)&=&1\label{norm1}\\ 
\int dy\; y\;f_N(y)&=&1  .\label{norm2}
\end{eqnarray}
The latter equation is remarkable; it states that
in deep inelastic scattering the nucleons act as if they carry all of the
$P^+$ of the nucleus  even though the mesonic fields are very prominent.

This is clearer if we re-interpret these sum rules in terms of a
probability $f_N(k^+)$ that a nucleon has a plus momentum
$k^+\equiv y\overline{M}$, with $f_N(k^+)\equiv A
f_N(y\overline{M})/\overline{M},$ so that \bea
&&\int dk^+\;f_N(k^+)=A,\label{baryonsr}\\
&&\int dk^+ \;k^+ 
\;f_N(k^+)= A\overline{M}=M_A\label{momsr}\eea
The momentum sum rule (\ref{momsr}) shows the total plus momentum carried
by the nucleons (as seen in deep inelastic scattering)
is also the total momentum carried by the nucleus.

The main result of this is that the nuclear structure
function is given by \eqn{good}) with the function $f_N(y)$ obtained in
\eqn{shift}).
This tells us that,
despite the fact that there is considerable binding energy, there is no
EMC binding effect.
Indeed,  $F_{2A}$ depends on the Fermi momentum but
does not depend on the effective mass $M^*$.

The quantity measured in deep inelastic scattering is the ratio
defined by
\begin{equation}
\label{rox} R(x)=\frac{F_{2A}(x_A)}{A F_{2N}(x)}.
\end{equation}
A numerical study of this expression
using,  five different relativistic models is presented below.
First, we  emphasize the qualitative features.
Since the
width of  $f_N$ is given by the small quantity $k_F/\overline{M}$  it
is a very  narrow
function. In this case,
one may evaluate the integrand of \eqn{good})
by expanding $F_{2N}(x/y)$ in a Taylor
series about $y=1$\cite {fsplb} to find that
\be
R(x)={F_{2N}(x_A)\over F_{2N}(x)}+
{k_F^2\over 10 \overline{M}^2F_{2N}(x)}\left(2x_AF'_{2N}(x_A)+
  x_A^2F''_{2N}(x_A)\right),
\label{expand1}\ee
which shows that the only effect of the binding energy occurs in the small
difference between $x_A$ and $x$ which depends only on the small average
binding energy.
 Note that a term proportional to
$F_{2N}'(x)$ (but not proportional to the small parameter
${k_F^2\over \overline{M}^2})$) vanishes because one is expanding about $y=1$
and using
 the baryon and momentum  sum rules
Eqs.~(\ref{baryonsr},\ref{momsr}).
We may further approximate $R(x)$ by expanding the first term about the value
$x_A=x$ 
($x_A=1.02\; x$ for nuclear matter,  and  $x_A\le1.01x\;$ for
finite nuclei). Thus \be
R(x)=
1+{\langle\epsilon\rangle\over \overline{M}}{F'_{2N}(x)\over F_{2N}(x) }
+{k_F^2\over 10 \overline{M}^2F_{2N}(x)}\left(2x_AF'_{2N}(x_A)+
  x_A^2F''_{2N}(x_A)\right),
\label{expand}\ee
where $\langle\epsilon\rangle$ is the binding energy per nucleon
(16 MeV for infinite
nuclear matter and $\le8$ MeV for finite nuclei).
This shows that, as long as the Hugenholtz-van Hove theorem is applicable,
the only binding effect is due to the binding energy per nucleon. One is not
allowed to use the separation 
energy which is much larger. The use of \eqn{expand}) cannot lead
to a large enough depletion of $R(x)$ \cite{birse}  to resemble
the extrapolated  data for nuclear matter\cite{Sick:1992pw}.
\begin{table}
\caption{Summary of the Models-Taking the sum of $ g_v{V}^{\,+}/\overline{M}$
  and $E_F^*/\overline{M}$  shows that each model satisfies the Hugenholtz-van
  Hove theorem, Eq.~(\protect\ref{hvh}).
  } \label{tab:summary}
\begin{tabular}{ccccccc}
Model & $\qquad g_v{V}^{\,+}\qquad$ & $\quad
g_v{V}^{\,+}/\overline{M}\quad$ & \quad$E_F^*/\overline{M}$ &
$\quad M^{*}/M\quad$ & $\quad k_{F}$
(fm$^{-1}$) &  \quad$P^+_N/P^+_A$\\
\hline
W & $g_{v}^{2}\rho_{B}/m_{v}^{2}$ & 0.355 & 0.645 & 0.56 & 1.42  & 0.65\\
NVW & $2G\rho_{B}$ & 0.355 & 0.645 & 0.56 & 1.42  & 1\\
ZM & $g_{v}^{2}\rho_{B}/m_{v}^{2}$ & 0.079 & 0.921 & 0.85 & 1.42 & 0.92\\
NVZM & $2G\rho_{B}$ & 0.079 & 0.921 & 0.85 & 1.42 & 1\\
RF & $2G_{RF}\rho_{B}$ & 0.194 & 0.806 & 0.73 & 1.31 & 1\\
\end{tabular}
\end{table}
\begin{figure}
\centering
\epsffile{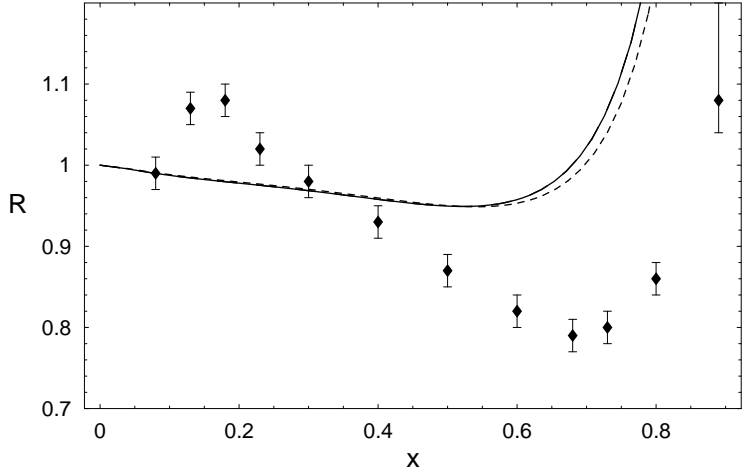}
\caption{$R(x)$ vs. $x$ for W, NVW, ZM, NVZM (solid line) and RF
(dashed line).
The data shown here are from the extrapolation of
Ref.~\protect\cite{Sick:1992pw}}
\label{fig:modelplots}
\end{figure}

The relevant
parameters of these models are displayed in Table~~\ref{tab:summary}.
The qualitative features discussed above are prominent in the numerical
calculations displayed in
Fig.~\ref{fig:modelplots} in  which  the ratio $R(x)$ of \eqn{rox})
is presented for  five different relativistic models. The relevant
parameters of these models are displayed in Table~~\ref{tab:summary}.
We note that four of the models have identical results, with only the
Rusnak-Furnstahl model (with its chosen different value of $k_F$)  differing
only very slightly.
These results are obtained
using a simple early
parameterization \cite{old} of the free nucleon structure function
\be
F_{2N}(x)=0.58\sqrt{x}(1-x)^{2.8}+0.33\sqrt{x}(1-x)^{3.8}+0.49(1-x)^{8},\ee
but the essential features of the curves are independent of the free nucleon
structure function.

\section{The Canonical Formalism--Walecka Model}
\label{walecka}
We  illustrate the canonical light front formalism using the
Walecka model as our first example. Some
of this  has been published previously\cite{Miller:1997cr},
but our purposes here are  to set up
and  illustrate
the formalism necessary to go beyond the mean field approximation,
provide an explicit
example of the general results presented in Section~II, study the nuclear
structure origins
of the nuclear momentum content, and show
 explicitly that the Hugenholtz-van Hove theorem is satisfied.

The Walecka model employs a
Lagrangian containing fields for  nucleons ($\psi'$), scalar mesons $\phi$
and vector mesons $V^\mu$: 
\begin{eqnarray}
{\cal L}_{W} & = &
\frac{1}{2}(\partial^{\mu}\phi\partial_{\mu}\phi-m_{s}^{2}\phi^{2}) 
-\frac{1}{4}V^{\mu\nu}V_{\mu\nu}+\frac{1}{2}m_{v}^{2}V^{\mu}V_{\mu} 
+\overline{\psi}'\left(\gamma^{\mu}(i\partial_{\mu}-g_{v}V_{\mu})-M-g_{s}\phi\right)
\psi',\end{eqnarray}
with the field equations:
\begin{eqnarray}
\label{phieom}
(\partial_{\mu}\partial^{\mu}+m_{s}^{2})\phi=-g_{s}\overline{\psi'}\psi'\\
\label{veq}
\partial_\mu V^{\mu\nu}+m_v^2V^\nu=g_v\overline{\psi}'\gamma^\nu\psi'\\
\label{diraceq1}
\gamma^{\mu}(i\partial_{\mu}-V_{\mu})\psi'=(M+g_{s}\phi)\psi'
.\end{eqnarray}
The symmetric canonical energy-momentum tensor is given
by\cite{soper,yan12,yan34}
\be T^{\mu\nu}=-g^{\mu\nu}{\cal L} +V^{\alpha\mu}V^{\beta\nu}g_{\beta\alpha}
+\partial^{\mu}\phi\partial^{\nu}\phi
+\frac{1}{2}\overline{\psi'}(\gamma^{\mu}(i\partial^{\nu}-g_vV^\nu)
+\gamma^{\nu}(i\partial^{\mu}-g_vV^\mu))\psi'.\label{tmunu}\ee
\subsection{ Mean Field Approximation for Infinite Nuclear Matter}
We follow the MFA  \cite{Walecka:1974qa} in assuming  that the
sources are sufficiently strong so that the resulting  large
numbers of mesons can be treated in  a classical manner in which
source operators  are replaced by their expectation values in the
nuclear ground state. Furthermore, for a system of infinite
volume, all positions and directions (in the nuclear rest frame)
are  equivalent. In this case $\phi$ and $V^0$ are constants and
$\mathbf{V}=0$. The approximate mesonic equations of motion become
\bea \phi & = &
-\frac{g_{s}}{m_{s}^{2}}\langle\overline{\psi'}\psi'\rangle
=-\frac{g_{s}}{m_{s}^{2}}\rho_s\label{phimf}\\
V^0&=&V^\pm=\frac{g_{v}}{m_{v}^{2}}\langle{\psi'}^\dagger\psi'\rangle
=\frac{g_{v}}{m_{v}^{2}}\rho_B,
\label{v0mf}\eea
in which the brackets are used as an abbreviation for taking the ground
state matrix element, and
\bea  \rho_B=2k_F^3/3\pi^2.\label{rhob}\eea
The nucleon, though described in terms of four-component spinors, consists of
only
two independent fields. The independent and dependent degrees of freedom are
defined by the projection operators:
$\Lambda_\pm\equiv {1\over 2}\gamma^0\gamma^\pm,\; \psi'_\pm\equiv
\Lambda_\pm \psi',$ 
with
$\psi'_+$ chosen as the independent field. One more step is necessary
because the resulting equation for $\psi'_-$  depends on $V^+$ in a complicated
manner.
It is traditional in light front dynamics of mass-less vector bosons
to remove the effects of the
term $V^+$, by working in a gauge with $V^+=0$. Here
we use the Soper-Yan transformation\cite{soper,yan34}:
\be
\psi'\equiv e^{-ig_v\Lambda}\psi,\qquad \partial^+\Lambda=V^+,\label{tsy}\ee
which allows a simple equation for  $\psi'_-$ in terms of  $\psi'_+$
to proceed in a
satisfactory manner, but which also causes the loss  of  manifest covariance.
With this transformation
the final version of the nucleon field equation becomes
\bea
(i\partial^--g_vV^-)\psi_+=(\bbox{\alpha}_\perp\cdot{\bf
  p}_\perp+\beta(M+g_s\phi))\psi_-\nonumber\\
i\partial^+\psi_-=(\bbox{\alpha}_\perp\cdot{\bf p}_\perp
+\beta(M+g_s\phi))\psi_+\eea
within the mean field approximation (in which $\partial^-\Lambda=0$)
for infinite nuclear matter.
The nucleon mode functions are plane waves so $ \psi\sim e^{ik\cdot x}$ and
\bea 
(i\partial^--g_vV^-)\psi_+={k_\perp^2+{M^*}^2\over k^+}\psi_+,\label{eig}\eea
where $
M^*\equiv M+g_s\phi.$

The relevant components of the energy-momentum tensor in the mean field
approximation MFA are obtained by using the constant meson fields of
Eq.~(\ref{phimf}) and (\ref{v0mf}) in Eq.~(\ref{tmunu}) to obtain
\begin{eqnarray}
\label{t++wal}
 T^{++}_{MFA} & = & m_{v}^{2}V_{0}^{2}+2\psi^{\dag}_{+}i\partial^{+}\psi_{+}\\
\label{t+-wal}
 T^{+-}_{MFA} & = & m_{s}^{2}\phi^{2} + 2\psi^{\dag}_{+}(i
\partial^{-}-g_{v}{V}^{-})\psi_{+}
\end{eqnarray}
and so $P^{+}$ and $P^{-}$ are given by
\be P^\pm=\langle T^{+\pm}_{MFA}\rangle\Omega,\label{pdefs}\ee
where
$\Omega $ is the volume of the system,  taken as  infinite at
the end of the calculation ($A,\Omega\to\infty$ with $A/\Omega$ finite).
The  evaluation of these expectation values 
yields
\bea
\label{p-wal} \frac{P}{\Omega}^{-} & = &
m_{s}^{2}\phi^{2}+\frac{4}{(2\pi)^{3}}\int_{F}
d^{2}k_{\bot}dk^{+}\frac{k_{\bot}^{2}+(M^{*})^{2}}{k^{+}}\\
\label{p+wal} \frac{P}{\Omega}^{+} & = & m_{v}^{2}V_{0}^{2}+
\frac{4}{(2\pi)^{3}}\int_{F} d^{2}k_{\bot}dk^{+}k^{+}.\eea It is
necessary to define the Fermi sea within the present context.
Although  we do not have manifest rotational invariance here, this
invariance is restored in the results if we define the component
$k^3$ implicitly  through
 \eqn{k3}). Then
 \bea
 \int_F d^3k\cdots&\equiv&\int d^3k\theta(k_F-k)\cdots,\label{theta}\eea
 and geometry leads to
 \bea
\int_{F} d^2k_{\perp}dk^{+}\cdots&\equiv&\int d^{2}k_{\bot}dk^{+}
\theta\left(k_F^2-k_\perp^2-(k^+-E_F^*)^2\right)\cdots.\eea Using
Eqs.~(\ref{p-wal}-\ref{theta}) leads to the results that the value
of the
 energy  of the system in the  rest-frame,
\bea E_A\equiv {1\over2}(P^++P^-),\label{ea}\eea
is the same as in the usual treatment of
the Walecka model, as shown below.
The only remaining task is to determine        the Fermi momentum,
$k_F$. This is done by using the minimization
\bea
\left(\frac{\partial (E_A/A)}{\partial k_F}\right)_\Omega=0,\label{partial}\\
E_A(k_F)=M_A.
\eea

Carrying out the differentiation leads to an equation which
is equivalent to setting $P^+=P^-$, which must occur for a system in its
rest frame with $P^3=0$. Since rotational
invariance is maintained in the solution $P^{1,2,3}=0$, and therefore
the pressure $P=1/3\sum_{i=1,3}P^i$ vanishes. Thus the equation
$P^+=P^-$  is also the light front equivalent of
setting the pressure $P$ to 0.
Note also that one may explicitly carry out the differentiation to find
that \be E_A/A=M_A/A=E^*_F+g_vV^0=E_F\label{hvh1}\ee
which is the Hugenholtz-van Hove theorem\cite{HvH}.

The above paragraph is serves as an outline of the derivation of
the Hugenholtz-van Hove theorem, but we also
provide an explicit proof. 
First, use the transformation (\ref{k3}) to obtain the results
\bea
\label{p-walk3} \frac{P}{\Omega}^{-} & = &
m_{s}^{2}\phi^{2}+\frac{4}{(2\pi)^{3}}\int_{F}
d^{3}k (E^*(k)-{1\over3}\bbox{k}\cdot\bbox{k})\\
\label{p+walk3} \frac{P}{\Omega}^{+} & = & m_{v}^{2}V_{0}^{2}+
\frac{4}{(2\pi)^{3}}\int_{F} {d^{3}k}(E^*(k)+{1\over3E^*(k)}\bbox{k}\cdot\bbox{k})
\\{E_A\over \Omega}&=& {1\over 2} m_s^2\phi^2 +
{1\over 2}m_v^2 V_0^2
+{4\over (2\pi)^3}\int_F d^3k E^*(k)
\label{Ei}
\eea

Next carry out the differentiation in \eqn{partial}),
using $A=\rho_B\Omega$
to obtain
\begin{equation}
{\partial E_A\over \partial k_F}=3{E_A\over k_F}.
\end{equation}
The term ${\partial E_A\over \partial k_F}$
is obtained
by first eliminating all derivatives with respect to
$\phi$ using the feature that
setting ${\partial E_A\over \partial \phi}$
 to zero reproduces the field
equation for $\phi$.  Then one 
 uses
the field equation for the vector meson (\ref{v0mf}).
 The result
 is\begin{equation}
{4\over (2\pi)^3}{4\pi\over 3}k_f^3 E_F^*={m_s^2\over 2}\phi^2-{m_v^2\over 2}
V_0^2+
{4\over (2\pi)^3}\int_F d^3k\;E^*(k),
\label{mide}
\end{equation}
This is a transcendental equation which determines $k_F$, so
that the calculation of $E_A$ is complete.
With the self-consistent Eqs.~(\ref{phimf},\ref{v0mf},\ref{Ei})
one obtains an average binding energy of 15.75 MeV with
a Fermi momentum of $k_F=1.42$ fm$^{-1}$ using the parameters:
$ \frac{g_{v}^{2}}{m_{v}^{2}}M^{2}  =  195.9,
 \frac{g_{s}^{2}}{m_{s}^{2}}M^{2}  =  267.1,$ which
 corresponds to $ g_vV^-=323$ MeV, and 
 $M^*/M=0.56$. 
These parameters are the same as in the original Walecka model.

The relation
$P^+=P^-$ (which must hold for a system in its rest  frame) also emerges as
a result of this minimization. To see this,   rewrite
the left hand side of
Eq.~(\ref{mide}) as
\begin{equation}
{4\over (2\pi)^3}{4\pi\over 3}k_f^3 E_F^*=
{4\over (2\pi)^3}
\int_F d^3k\left( E^*(k)+
{\bbox{k}\cdot\bbox{k}\over 3 E^*(k)}\right).
\end{equation}
Using this in Eq.~(\ref{mide}) leads to
\begin{equation}
{m_s^2\over 2}\phi^2-{m_v^2\over 2}
V_0^2= {4\over (2\pi)^3}\int_F
d^3k{\bbox{k}\cdot\bbox{k}\over 3 E^*(k)},\label{pb}
\end{equation}
which is what one obtains by setting $P^+=P^-$ using
Eqs.~(\ref{p-walk3}) and (\ref{p+walk3}).

We now have the tools at hand to prove the Hugenholtz-van Hove theorem.
Simply use  \eqn{mide}) to remove  the integral appearing in
\eqn{Ei}) and obtain
\bea{E_A\over\Omega}=m_v^2V_0^2+\rho_BE_F^*.\eea
Then the use of the field equation (\ref{v0mf}) yields
\be {E_A\over\rho_B\Omega}={E_A\over A}=g_vV^0+E_F^*=E_F,\label{hvhnum}\ee
which is the desired result.
This is a remarkable result. The original version of the theorem was proved using
only the assumption that nucleons are the only degrees of freedom. Here, the
mesons are important, yet the theorem still  holds\cite{nada}.

\subsection {Nuclear Plus-Momentum Content}
Now we relate  the role of
the plus component of the momentum seen here with that of
Section~\ref{ratiofunction}.
The nucleonic contribution to the nuclear
plus momentum from \eqn{p+wal}) is 
\begin{equation}
{P^+_N\over A}={4\over\rho_B (2\pi)^3}\int_F d^2k_\perp dk^+ k^+,\label{nuc}
\end{equation}
which is also obtainable directly from taking the nuclear expectation
value of nucleon plus momentum operator.

The large vector potential and small value of $M^*$ are associated with the
startling result
 that   only 65\% of the plus momentum of the nucleus is carried by
nucleons, and that 35\%
is carried by vector mesons.
 It was previously argued\cite{Miller:1997xh,Miller:1997cr}
that this would produce a disastrously large
decrease in the nuclear deep inelastic structure function.
As shown above, that does not occur, because the function $f_N(y)$ peaks at
$y=1$.

\begin{figure}
\centering
\epsffile{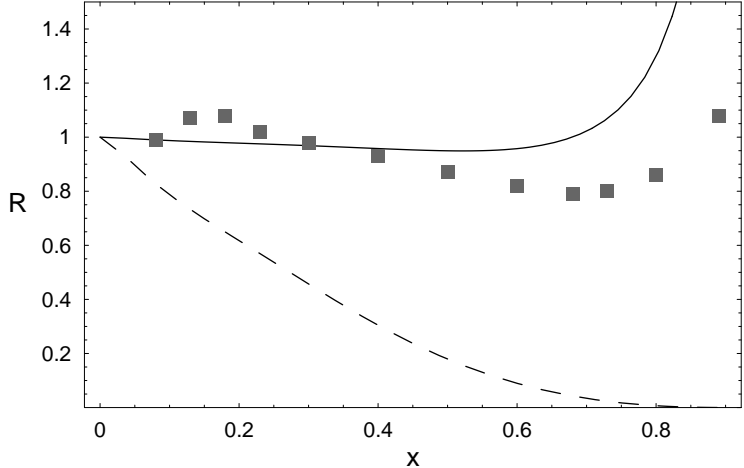}
\caption{$R(x)$ vs. $x$ for the Walecka model  using $f_N(y)$ (solid line)
  and $f(y)$ in \protect\eqn{deeper})
(dashed line).
The data shown here are from the extrapolation of
Ref.~\protect\cite{Sick:1992pw}}

\label{fig:compare} 
\end{figure}

We therefore need to understand how it is that
the nucleons can carry 65\% of the momentum here
all of the momentum as stated in Section~\ref{ratiofunction}.
To do this, use \eqn{nuc}) to define
a probability $f(k^+)$ that a nucleon carries a plus momentum
$k^+$:
\begin{equation}P^+_N/A=\int
dk^+\;k^+ f(k^+),
\end{equation}
with
\bea
f(k^+)={4\over\rho_B (2\pi)^3}\int_F d^2k_\perp=
{4\over\rho_B (2\pi)^3}\int_F d^2k_\perp dp^+\delta(k^+-p^+).\label{fdef}
\eea
It is useful to again
obtain a dimensionless distribution
function $f(y)$  by
replacing $k^+$ by the dimensionless variable $y$ using
$y\equiv {k^+\over \overline{M}},f(y)\equiv\overline{M}f(k^+).$
 Then one finds \begin{equation}
f(y)={3\over 4} {\overline{M}^3\over k_F^3}\theta(y^+-y)\theta(y-y^-)\left[
{k_F^2\over \overline{M}^2}-({E_F^*\over \overline{M}}-y)^2\right], \label{fy}
\end{equation}
where
$y^\pm\equiv {E_F^*\pm k_F\over \overline{M}}.$
This function peaks at $y=E_F^*/\overline{M}=0.65$ for the Walecka model, and
the average value of $y$ is also 0.65. Its use in \eqn{deeper}) would indeed
lead to a disaster. Indeed,
our previous work assumed that  Eq.~(\ref{deep}) was appropriate. In that case,
the computed ratio
${F_{2A}(x)\over A}$ was  dramatically  smaller than
$F_{2N}$. This disastrous result can be understood from the following logic.
A reasonable first approximation to the integral   Eq.~(\ref{deep})
can be obtained by using $f(y)\approx \delta (y-
{E_F^*\over\overline{M}})$
which satisfies the baryon sum rule and
corresponds to an average value of $y=0.65$. Then
${F_{2A}(x)\over A}$ vanishes for $x>0.65$
and the ratio to the free structure function goes to zero
in huge contradiction with experiment, which shows depletions no larger than
20\% for the heaviest nuclear targets. This result is illustrated with the
numerical calculation shown in Fig.~\ref{fig:compare}.

However,
the correct quantity to use in deep inelastic scattering
is $f_N(y)$, which emerges from
a manifestly covariant treatment. To see the connection between $f(y) $ and
$f_N(y)$, it is first helpful to compare directly Eqs.~(\ref{shift})
and (\ref{fy}). This comparison yields:
\be f(y) =f_N(y+g_vV^+/\overline{M}),\label{both}\ee
 using  the Hugenholtz-van Hove theorem~\eqn{hvh}).
 The  correct nucleon distribution function is
one that is shifted by
the vector potential. This simple relation suggests that there is a simple
interpretation of
the difference between $f_N(y)$ and $f(y)$ in terms of a phase difference.
Indeed, the difference arises from the Soper-Yan transformation
(\ref{tsy}) which relates the fields $\psi$ and $\psi'$. Thus we have two forms
of the plus momentum density, $T^{++}_{MFA}$:
\begin{eqnarray}
\label{t++walr}
 T^{++}_{MFA} & = & m_{v}^{2}V_{0}^{2}+2\psi^{\dag}_{+}i\partial^{+}\psi_{+}
 =2 \psi'{^{\dag}}_{+}\left(i\partial^{+}-g_vV^+\right)
 \psi_{+}'.\label{2++}\eea
 In the second form, a nucleon operator carries all of the plus momentum,
 the correct single nucleon plus momentum operator is the canonical
 conjugate momentum which  is shifted by the term $g_vV^+$. The second form is
 the appropriate one as it is related to the original covariant Lagrangian.
 The term $yf_N(y)$ is obtained from the
expectation value of the nucleon plus-momentum operator \
$\psi'^\dagger_+ (i\partial^+-g_vV^+)\psi_+',$ while $yf(y)$ is
obtained from $\psi^\dagger_+ i\partial ^+\psi_+$ with   $f_N(y)$
as the distribution function which emerges from a covariant
treatment.

\section {Four More Models in Mean Field Approximation}
\label{four}

The Walecka model, evaluated in MFA, was known to have some phenomenological
troubles. The compressibility is  too large, and the very small effective
mass (shown here to be irrelevant for deep inelastic scattering) does enter
into quasi-elastic scattering, in which it is a straightforward matter
to show that the cross section is given by an integral over the distribution,
 $f(y)$, and not $f_N(y)$.
The reason for this  is that in the MFA the struck
nucleon feels the same vector
and scalar potentials as a bound nucleon. Hence it was of interest to
improve the Lagrangian. This has been done in a variety of ways.
We consider four other models here, mainly to show that the same
function $f_N(y)$ emerges from each one and that each one satisfies the
Hugenholtz-van Hove theorem. The validity of this theorem is a signal that
the nucleons carry all of the plus-momentum,  just as for the Walecka model,
so that there can be no significant binding effect.

All of the  models have essentially the same saturation properties, but
each is
distinguished by using a different mechanism to  reduce the putative  amount
of momentum carried by the vector mesons. Here we simply define the
models and summarize the results. The details of the solution are presented in
the Appendix. The important parameters of each model are displayed in
Table~\ref{tab:summary}.

The No Vector Wale{c}ka (NVW) model
is defined by the elimination of the vector meson field in favor of
a point coupling interaction of the form $Gj^\mu j_\mu$ with
$j_{\mu}\equiv \overline{\psi'}\gamma_{\mu}\psi'$.
In this case the nucleons
carry all of the momentum.
The Zimanyi-Moszkowski (ZM) model\cite{Zimanyi:1990np}
is defined by using a
rescaled derivative coupling
interaction in which the scalar coupling in the Lagrangian is given
by the term $
-\overline{\psi}'{M}/({1-\frac{g_{s}\phi}{M}})\psi'.$
This model is known to involve a larger effective mass
 and smaller vector field
than the Walecka model. The Appendix shows that, in this model,
the nucleons carry about 92\% of the total plus momentum. The
 No Vector Zimanyi-Moszkowski (NVZM) model is defined by starting with the
Zimanyi-Moszkowski Lagrangian
and then
removing the  vector mesons in favor of a  current-current interaction as in
the NVW model. Again  nucleons carry all of the plus-momentum.
The
Rusnak-Furnstahl (RF)
point coupling model contains no explicit meson fields, and
the interactions included via a variety of non-linear couplings.
The parameters are given in Ref.~\cite{Rusnak:1997dj}, and arguments for
their naturalness in terms of effective field theory have also been presented
\cite{Furnstahl:2000rm}.
The nucleons
carry all of the plus-momentum and
the numerical value of the effective mass is
$M^{*}=0.73M$,  significantly higher
than
that of the Walecka model.

Each of the models summarized in Table~\ref{tab:summary}
has essentially the same saturation properties even though
the values of $M^*$ and $E_F^*$  display huge variations.
Note that each model has a nucleon mode equation (\ref{eig}),
(\ref{dirsol}),  (\ref{dirsolzim}),
(\ref{dirsolzimnv}), and (\ref{dirsolfurn}), and these are  summarized by the
single unifying expression:
\begin{equation}
\label{dirsolratio} (i\partial^{-}-g_v{V}^{\,+})\psi_{+}
=\frac{k_{\bot}^{2}+(M^{*})^{2}}{k^{+}} \psi_{+}
.\end{equation}
The numerical values of $g_v{V}^{\,+} $ are listed
 along with other relevant information regarding  the
five models in Table \ref{tab:summary}.

\section{Beyond the Mean Field Approximation}
\label{beyond}
It is worthwhile to discuss the generality of the result that
there is no significant  binding effect. Consider any model, such that of
Ref.~\cite{Pieper:2001mp}
 in which mesonic fields are not
explicit components of the nuclear Fock state wave function.
For example, one may eliminate the mesonic degrees of freedom in favor of
 two- and three- nucleon interactions without maintaining the mesonic presence
in the nuclear Fock state wave function.
Such models, correctly evaluated,  obey
the Hugenholtz-van Hove theorem, Eq.~(\ref{hvh}).
The validity of this
theorem is a signal that nucleons carry all  of the plus momentum, so that the
baryon and momentum sum rules Eqs.~(\ref{norm1}-\ref{momsr}) are satisfied.
This means that for equilibrium the following conditions hold
\bea P^\pm&=&P^\pm_N\\
P^+&=&P^-.\eea
These two equations (the first is defined by the model, the second by
stability)  may be thought of as the light front version of the
Hugenholtz-van Hove theorem.
Therefore one may again apply the
analysis of Refs.~\cite{fs2,fsplb} and expand $f_N(y)$, appearing in the
integral of Eq.~(\ref{good}) about its peak value of
unity. It is generally  sufficient to keep only
three  terms. Thus one finds
\bea F_{2A}(x)&=&F_{2N}(x_A) +\gamma \left(2x_AF'_{2N}(x_A)+
  x_A^2F''_{2N}(x_A)\right)\label{exp}\\
\gamma&\equiv&\int dy\;(y-1)^2 f_N(y).\eea
The coefficient $\gamma$ is larger than the term proportion  to $k_F^2$
of \eqn{expand}) because the effects of correlations
extend the width of the distribution $f_N(y)$, but it multiplies a term which
is positive in the valence quark region. Thus, once again we see that
the only binding effect appears in the presence of the variable $x_A$ which is
only slightly larger than $x$, see \eqn{xa}.
This effect is too small to reproduce the data;
there is essentially
no EMC binding effect. The result (\ref{exp})
is                very similar to Eq.~(\ref{expand}) in that
 the   term, usually associated
with the binding effect,  proportional to
${\partial\over \partial y} F_{2N}(x_A/y)\vert_{y=1}$
vanishes because of the second sum rule of Eq.~(\ref{hvh}). We stress that
any model in which nucleons are the only degrees of freedom must obey the
momentum sum rule, as expressed in either \eqn{norm2}) or (\ref{momsr}).
Thus \eqn{exp}) will emerge and the model will not have a sufficiently large
binding effect to explain the nuclear deep inelastic scattering data
at large  values of $x$.

Can the conventional meson-nucleon picture of nuclear structure (which ignores
off-shell effects) be used to
reproduce the nuclear deep inelastic scattering data in the valence quark
region of Bjorken $x$? The only way to get a binding effect is to compute the
nuclear ground state wave function in such a manner as to obtain the
mesonic $P_m^\pm$ and nucleonic contributions: 
\bea P^+=P_N^+ + P^+_m,\label{srp}\\
 P^-=P_N^- + P^-_m,\label{srm}\eea
in which the meson content $P^\pm_m$ is treated  explicitly.
 In general, $P_m^\pm$ consists of terms
arising from any of the exchanged  mesons which are responsible for the
nuclear force:\bea P_m^\pm =P^\pm_\pi + P^\pm_\omega  + P^\pm_\sigma+ \cdots.
\eea
The equation for $P^+$ was used long ago\cite{Berger:1984jk}, in which nucleons
and pions contributed to the total, and with momentum
conservation presented as the justification for the equation. It is useful
to realize that the use of the energy-momentum tensor provides
a general basis for this sum rule. For example, in the work of
Refs.~\cite{Miller:1997cr,Miller:1999ap} the use of a chiral  Lagrangian,
containing isoscalar vector mesons, $V^\mu, V^{\mu\nu}=\partial^\mu V^\nu-
\partial^\nu V^\mu$, scalar mesons $\phi$ and pions
$\bbox{\pi}$, and standard manipulations give
\begin{eqnarray}
T^{++}&=&V^{ik}V^{ik}
+m_v^2V^+ V^+
+\bar\psi\gamma^+ i\partial^+ \psi\nonumber\\
&+&\partial^+\phi\partial^+\phi
+\partial^+\bbox{\pi}\cdot \partial^+ \bbox{\pi}
+\bbox{\pi}\cdot\partial^+\bbox{\pi}
{\bbox{\pi}\cdot\partial^+\bbox{\pi}\over\pi^2}(1-{f^2\over \pi^2}
\mbox{sin}^2{\pi\over f}), \label{tpp}
\end{eqnarray}
with $P^+$ given by
\be P^+=\langle T^{++}\rangle \Omega,\ee
with the brackets denoting a ground state matrix element.
The point is that each separate term corresponds to a term in either $P_N^+$ or
$P_m^+$, and therefore identifiable as a term in the sum rule (\ref{srp}).
However, the field equations provide relations between all of the
fields. Thus,  one can not get a reasonable result
for $P^+$ by considering only one of the terms which contribute.

The pressure balance condition $P^+=P^-=A\overline{M}$,
must hold for a stable solution so that one finds
\be P_N^+-P^-_N=P_m^--P_m^+.\label{imb}\ee
Thus the condition needed to prove the
 the Hugenholtz-van Hove theorem (that $P^+_N=P^-_N=A\overline{M}$)
is not obtained. We know that $P_N^+<M_A$ because all of the contributions
to $P^+_m$ are positive definite.
One may therefore define a positive
quantity $\epsilon$ via the deviation:
\bea \epsilon \equiv {P^+_m\over M_A},\label{eps}\eea
so that \bea
\int dy\; yf_N(y)=1-\epsilon.\label{dev}\eea
Thus Eqs.~(\ref{imb},\ref{eps}) can be thought of a generalization of the
Hugenholtz-van Hove theorem which is equivalent to the momentum sum rule.
With this new feature, the application of the
expansion procedure to the integral of \eqn{deeper})
yields  a term proportional to $\epsilon$:
\bea F_{2A}(x)&=&F_{2N}(x_A) +
\epsilon x_AF'_{2N}(x_A)
+\gamma \left(2x_AF'_{2N}(x_A)+
  x_A^2F''_{2N}(x_A)\label{cbe}\right).\eea
Equation~(\ref{cbe}) corresponds to the usual binding effect which
now is present. However, one needs fairly large values
$\epsilon\sim 0.05$ to reproduce the deep inelastic data for
Iron\cite{fs2}, and $\epsilon\sim 0.07$  to reproduce the nuclear
matter extrapolation shown in Fig.~\ref{fig:modelplots}. Early
calculations\cite{ET,Berger:1984jk} in which pions are allowed to
carry such a momentum fraction have another
consequence\cite{dyth}:
 an enhanced nuclear anti-quark content which turned out to be in
contrast with the results of the nuclear Drell-Yan experiment\cite{dyexp}.
A more recent light-front
calculation\cite{Miller:1999tp,Miller:1999ap}, which included the
effects of nucleon-nucleon correlations as well as those of an
explicit meson Fock space, finds that pions carry about 2\% of the
nuclear plus momentum. While this value is consistent with the
Drell-Yan experiment\cite{Miller:2001yf}, using $\epsilon=0.02$
would provide  too small a reduction in the nuclear structure
function.  It is possible that other mesons could supply
significant contributions to $\epsilon$, and it is necessary to
investigate this possibility. However, consistency with the
Drell-Yan experiment must be maintained. While  it seems unlikely
that a careful calculation will be consistent with both the deep
inelastic and Drell-Yan data, we cannot rule that out now.

The  analysis
of the previous
paragraph is  similar to the early one of Refs.~\cite{fs2} and \cite{Berger:1984jk}. The
main difference occurs in the present ability\cite{Miller:2000kv}
to compute the nuclear binding energy
in terms of $P_N^\pm$ and $P^\pm_m$.

It is also worthwhile commenting on the modern calculations\cite{Boffi}
which use
the nucleon spectral function $S(p)$ to   compute
the quantity $f_N(y)$.
It is necessary to  obtain \eqn{dev}) with a significantly large value
of $\epsilon$ to achieve agreement with data.
However, a complete calculation should also obtain
the very same value of $\epsilon$ from \eqn{eps}). But models in which
the mesons are eliminated in favor of two and three nucleon potentials, forfeit
the ability to compute the value of $\epsilon $ directly.
It is possible to make a 
completely accurate
calculation of $f_N(y)$ which would  reproduce the correct value of
$\epsilon$, but  a computational  error
which is only a few percent in $f_N(y)$ corresponds to a huge percentage error
in  the small quantity $\epsilon$. Hence,
such calculations must be regarded as inconclusive. Even
if we take the calculations
at face value, the models ``are not completely satisfactory''. If mean field
models are used, nuclear binding
accounts for only 20\% of the observed effect \cite{kulagin}.Very large
separation energies (values of $\epsilon$), inconsistent with the mean
field calculations, are required\cite{fsplb,Li}
 to reproduce the
data. Calculations have been made including correlations, but the
summary of Refs.~\cite{ciofi,lex} made in the book\cite{Boffi} is:
``But for all nuclei considered the predicted deviation of the
ratio $R(x)$ (is) much smaller than the experimental one.'' This
statement applies also to  the work of \cite{vj}, if the
``off-shell nature of the nucleon'' is ignored. The inclusion of
off-shell effects by allowing the nucleon structure function to
depend on  the momentum of the nucleon in the nucleus (as well as
on $x/y$) can lead to a significantly improved description of  the
data\cite{gl,vj}. This agreement is consistent with the results of
the present work. Here we  consistently ignore off-shell effects,
categorizing these, along with a host of others, as interesting
effects. In any case, one would need to understand the
implications of analogous off-shell variations in operators used
in impulse approximation calculations for many nuclear reactions.
Furthermore, it is not clear to us that these formulations
\cite{gl,vj} provide nuclear structure functions which are
consistent with the baryon sum rule.
\section{Summary and Discussion}
\label{summary} The principal result is that relativistic mean
field models of nuclei, successful for many observables, do not
contain the binding effect needed to reproduce the depletion
observed by the EMC. The generality of this conclusion is related
to the use of the mean field approximation, consistent with the
Hugenholtz-van Hove theorem\cite{HvH}, which severely constrains
the nucleon distribution function $f_N(y)$. This theorem has the
further implication that any model in which the entire plus
momentum is carried by the nucleons, in the sense of
Eq.~(\ref{hvh}), also contains no binding effect. Thus including
nucleon-nucleon correlations (within a model containing only
nucleons in the Hamiltonian) cannot  reproduce  of the data. A
minimum feature  necessary to describe the data using conventional
meson-nucleon dynamics is that the mesonic components must
comprise an explicit part of the nuclear Fock state wave function,
and the mesons must carry a significant fraction of the nuclear
$P^+$. But there are severe constraints on the nuclear anti-quark
content\cite{dyth,dyexp} and these limit the flexibility of
mesonic models. Therefore, all of our present considerations are
consistent with the notion that some effect not contained within
the conventional framework is responsible for the EMC effect.
\section*{Acknowledgments}
We thank the USDOE for partial support of this work. We thank M. Burkardt,
A. Dieperink, U. Mosel, and
M. Strikman, and acknowledge the
UW INT for its support during the Spring 2001 program.
\appendix
\section {Four More Models}
\label{4app}
Four other
relativistic models are solved using light front dynamics in this section.
The techniques are the same as in
Section~\ref{walecka}, so our treatment will be
briefer than what appears above.
\subsection{The No Vector Wale{c}ka Model}
\label{novectora}

To reduce the effects of vector mesons it is natural to employ
 a Lagrangian that has only one meson field, a scalar field $\phi$. The
repulsion is supplied by a  repulsive vector point coupling. The
Lagrangian is
\begin{eqnarray}
          {\cal L}_{WNV} & = &
\frac{1}{2}(\partial^{\mu}\phi\partial_{\mu}\phi-m_{s}^{2}\phi^{2})
+\overline{\psi'}(i\partial\!\!\!\!\!\:/-M  -g_{s}\phi )\psi'
-G j^\mu\;j_\mu,
\end{eqnarray}
where
\be j_{\mu}\equiv \overline{\psi'}\gamma_{\mu}\psi'.
\ee
The resulting
equation of motion for the Dirac field is
\begin{eqnarray}
\gamma^{\mu}(i\partial_{\mu}-2Gj_{\mu})\psi'=(M+g_{s}\phi)\psi',
\end{eqnarray}
and the equation for the scalar field is again Eq.~(\ref{phieom})
The canonical energy-momentum tensor is given by
\be T^{\mu\nu}=-g^{\mu\nu}{\cal L}
+\partial^{\mu}\phi\partial^{\nu}\phi
+\frac{i}{2}
\overline{\psi'}(\gamma^{\mu}\partial^{\nu}+\gamma^{\nu}\partial^{\mu})\psi'.\ee

In the Mean Field Approximation (MFA) the equations of motion are given
by Eq.~(\ref{phimf}) for the scalar field and
\begin{eqnarray}
\gamma^{\mu}(i\partial_{\mu}-2G\langle j_{\mu}\rangle)\psi'& = &
(M+g_{s}\phi)\psi',
\end{eqnarray}
for the Dirac field.
The components of the energy-momentum tensor are given by
\begin{eqnarray}
T^{++}_{MFA} & = & i\overline{\psi'}\gamma^{+}\partial^{+}\psi'\\
T^{+-}_{MFA} & = & m_{s}^{2}\phi^{2}
-2\:\overline{\psi'}(\gamma^{\mu}(i\partial_{\mu}-G \langle
j_{\mu}\rangle)-M -g_{s}\phi )\psi'
+\frac{i}{2}\overline{\psi'}(\gamma^{+}\partial^{-}+\gamma^{-}\partial^{+})\psi'
\end{eqnarray}

We solve the Dirac equation as in Section~\ref{walecka}
using the transformation
\begin{equation}
\label{transform} \psi'=e^{-2iG\Lambda(x)} \psi,\quad 
\partial^{+}\Lambda=\langle j^{+}\rangle
\end{equation}
and define $\widetilde{j}^{\mu}=\langle
j^{\mu}\rangle-\partial^{\mu}\Lambda$ (note that
$\widetilde{j}^{+}=0$ by construction and $j^{i}=0$ in the rest
frame so that the only non-vanishing component in the MFA is
$\widetilde{j}^{-}=\rho_{B}$). The Dirac equations for $\psi_{+}$
and $\psi_{-}$ become
\begin{eqnarray}
\nonumber (i\partial^{-}-2G \widetilde{j}^{-})\psi_{+}& = &
(\alpha_{\bot} \cdot
p_{\bot}+\beta(M+g_{s}\phi))\psi_{-}\\
\nonumber i\partial^{+}\psi_{-} & = & (\alpha_{\bot} \cdot
p_{\bot}+\beta(M+g_{s}\phi))\psi_{+}
\end{eqnarray}
If we assume $\psi \sim e^{ik \cdot x}$, then we obtain
\begin{equation}
\label{dirsol}
(i\partial^{-}-2G\widetilde{j}^{-})\psi_{+}
=\frac{k_{\bot}^{2}+(M+g_{s}\phi)^{2}}{k^{+}}
\psi_{+}.
\end{equation}
 Returning to the energy momentum tensor,
which has also changed under the transformation Eq.
(\ref{transform}), we find
\begin{eqnarray}
 \label{t++}
 T^{++}_{MFA} & = & 2\psi^{\dag}_{+}(i\partial^{+}+2G\langle
 j^{+}\rangle)\psi_{+}\\
\label{t+-}
 T^{+-}_{MFA} & = & m_{s}^{2}\phi^{2} + 2\psi^{\dag}_{+}(i
\partial^{-}-2G\widetilde{j}^{-})\psi_{+}
.\end{eqnarray}
Using Eq. (\ref{dirsol}) in Eq. (\ref{t+-}) and the light front
4-momentum definition Eq. (\ref{pdefs})
we obtain
\begin{eqnarray}
\label{p-} \frac{P}{\Omega}^{-} & = &
m_{s}^{2}\phi^{2}+\frac{4}{(2\pi)^{3}}\int_{F}
d^{2}k_{\bot}dk^{+}\frac{k_{\bot}^{2}+(M^{*})^{2}}{k^{+}}\\
\label{p+} \frac{P}{\Omega}^{+} & = & \frac{4}{(2\pi)^{3}}\int_{F}
d^{2}k_{\bot}dk^{+}(k^{+}+2G\rho_{B})
\end{eqnarray}
The second term in Eq. (\ref{p+}) can be rewritten
\[\frac{4}{(2\pi)^{3}}\int_{F}
d^{2}k_{\bot}dk^{+}2G\rho_{B}
=\frac{8G\rho_{B}}{(2\pi)^{3}}\int_{F}d^{2}k_{\bot}dk^{+}=2G\rho_{B}^{2}\]
and the resulting equations are exactly those of the Wale{c}ka
model and we draw the correspondence (with $k_{F}=1.42\text{
fm}^{-1}$)
\bea 2GM^2\to {g_v^2\over m_v^2}M^2, \eea
which  means that the
saturation properties of this model are the same as those of the Walecka model.

An important difference between this model and the Wale{c}ka
model is that the extra term is \textit{not} due to vector mesons,
but part of the nucleon contribution to $P^{+}$. All of the plus
momentum is due to the nucleons and not the vector mesons.
It is apparent that this model is consistent with the Hugenholtz-van Hove
theorem. The values of $P^\pm$ are the same as those of the Walecka model for
all values of $k_F$.

\subsection{The Zimanyi-Moszkowski Model}
\label{zimanyia}

The rescaled derivative coupling Lagrangian given by Zimanyi and
Moszkowski \cite{Zimanyi:1990np} is known to have a  smaller vector potential
than
the Walecka model. This model is defined by the Lagrangian
\begin{eqnarray}
\nonumber \mathcal{L}_{ZM} & = &
\frac{1}{2}(\partial^{\mu}\phi\partial_{\mu}\phi-m_{s}^{2}\phi^{2})
-\frac{1}{4}V^{\mu\nu}V_{\mu\nu}+\frac{1}{2}m_{v}^{2}V^{\mu}V_{\mu}\\
&&
+\overline{\psi}\left(\gamma^{\mu}(i\partial_{\mu}-g_{v}V_{\mu})-\frac{M}{1-\frac{g_{s}\phi}{M}}\right)\psi.
\end{eqnarray}
Using the methods described in Section \ref{walecka}, one finds that
the eigenvalue equation  corresponding to Eq. (\ref{eig}) is given by
\begin{equation}
\label{dirsolzim} (i\partial^{-}-g_{v}{V}^{-})\psi_{+}
 =  \frac{k_{\bot}^{2}+(M^{*})^{2}}{k^{+}} \psi_{+}
\end{equation}
where
\bea M^{*}&=&
\frac{M}{1-\frac{g_{s}\phi}{M}}\label{mszm}\\
{V}^{-}&=&V_{0}=\frac{g_{v}\rho_{B}}{m_{v}^{2}}.
\eea
The field ${V}^{-}$ is transformed according to
 Eq.~(\ref{tsy}). The relevant components of the energy-momentum
tensor are
\begin{eqnarray}
\label{t++zim}
 T^{++}_{MFA} & = & m_{v}^{2}V_{0}^{2}+2\psi^{\dag}_{+}i\partial^{+}\psi_{+}\\
\label{t+-zim}
 T^{+-}_{MFA} & = & m_{s}^{2}\phi^{2} + 2\psi^{\dag}_{+}(i
\partial^{-}-g_{v}{V}^{-})\psi_{+}
\end{eqnarray}
and so $P^{+}$ and $P^{-}$ are given by
\begin{eqnarray}
\label{p-zim} \frac{P}{\Omega}^{-} & = &
m_{s}^{2}\phi^{2}+\frac{4}{(2\pi)^{3}}\int_{F}
d^{2}k_{\bot}dk^{+}\frac{k_{\bot}^{2}+(M^{*})^{2}}{k^{+}}\\
\label{p+zim} \frac{P}{\Omega}^{+} & = & m_{v}^{2}V_{0}^{2}+
\frac{4}{(2\pi)^{3}}\int_{F} d^{2}k_{\bot}dk^{+}k^{+}
\end{eqnarray}
which are superficially the same as the Wale{c}ka model,
but differ in that now $M^*$ is given by Eq. (\ref{mszm}).
The parameters of the model are obtained by minimizing the total energy
at
$k_{F} = 1.42 \text{ fm}^{-1}$, using the values
\begin{eqnarray}
\frac{g_{v}^{2}}{m_{v}^{2}}M^{2} & = & 43.2  \label{zvcoup}\\
\frac{g_{s}^{2}}{m_{s}^{2}}M^{2} & = & 140.4 \label{zscoup},\eea
which corresponds to \bea
M^{*}=0.85M.
\end{eqnarray}

The plus momentum may be decomposed using  $P^{+}=P^{+}_{m}+P^{+}_{N}$ (meson
part and a nucleon part)
\begin{eqnarray}
\frac{P^{+}_{m}}{A} & = &
\frac{m_{v}^{2}V_{0}^{2}}{\rho_{B}}  =  73\text{ MeV }\\
\frac{P^{+}_{N}}{A} & = &
\frac{1}{\rho_{B}}\frac{4}{(2\pi)^{3}}\int_{F}
d^{2}k_{\bot}dk^{+}k^{+}  =  850\text{ MeV }.
\end{eqnarray}
The nucleons carry about 92\% of the total plus momentum in this model.
Despite this, the Hugenholtz-van Hove theorem is satisfied because the
expressions for $P^\pm$ in terms of $V^0$ and $M^*$ are the same  as those of
the Walecka model. The only difference is the relation between $M^*$ and
$\phi$. However, that relation does not enter in the derivation of the
Hugenholtz-van Hove theorem presented in Section~III.
\subsection{A No Vector Zimanyi-Moszkowski Model}
\label{nvzimanyia}

The next  step is to modify the Zimanyi-Moszkowski Lagrangian by
removing the  vector mesons in favor of a  current-current interaction as in
Section \ref{novectora}
This
Lagrangian is
\begin{eqnarray}
\nonumber
\mathcal{L}_{NVZM} & = &
\frac{1}{2}(\partial^{\mu}\phi\partial_{\mu}\phi-m_{s}^{2}\phi^{2})
+\overline{\psi}\left(\gamma^{\mu}i\partial_{\mu}-\frac{M}{1-\frac{g_{s}\phi}{M}}\right)\psi\\
&&
-G\overline{\psi}\gamma^{\mu}\psi\overline{\psi}\gamma_{\mu}\psi.
\end{eqnarray}
The results follow exactly as those of Section \ref{novectora}.
Specifically, changing from a ``vector'' to ``no vector'' model
does not affect the minimization of the energy density and
therefore leaves the coupling constants (\ref{zvcoup}) and
(\ref{zscoup}) unchanged if we make the identification
$2GM^{2}\rightarrow (g_{v}^{2}/m_{v}^{2})M^{2}$. The operator
corresponding to Eq. (\ref{dirsol}) is
\begin{equation}
\label{dirsolzimnv} (i\partial^{-}-2G\widetilde{j}^{-})\psi_{+}
=\frac{k_{\bot}^{2}+(M^{*})^{2}}{k^{+}} \psi_{+}
\end{equation}
For this model $P^{-}$ is exactly Eq. (\ref{p-zim}) and we have
plus momentum
\begin{equation}
\label{p+znv} \frac{P}{\Omega}^{+}=\frac{4}{(2\pi)^{3}}\int_{F}
d^{2}k_{\bot}dk^{+}(k^{+}+2G\rho_{B})
\end{equation}

This model obeys the Hugenholtz-van Hove theorem because the values of
$P^\pm$ are the same as those of the Zimanyi-Moszkowski model for all values of
$k_F$.
\subsection{Rusnak-Furnstahl Point Coupling Model}
\label{furnstahla}

The point coupling Lagrangian of Ref.~\cite{Rusnak:1997dj}
is the modern version of the original Walecka which is connected to
QCD through  symmetry  and  naturalness. This model is defined  in
terms of the following densities:
 $j_{\mu}=\overline{\psi'}\gamma_{\mu}\psi'$,
$\rho_{s}=\overline{\psi'}\psi'$ and
$s_{\mu\nu}=\overline{\psi'}\sigma_{\mu\nu}\psi'$,
so that
\begin{eqnarray}
\nonumber \mathcal{L}_{RF} & = &
\overline{\psi'}(i\partial\!\!\!\!\!\:/-M )\psi'
-\rho_{s}^{2}(\kappa_{2}+\kappa_{3}\rho_{s}+\kappa_{4}\rho_{s}^{2})
-j_{\mu}j^{\mu}(\zeta_{2}+\eta_{1}\rho_{s}+\eta_{2}\rho_{s}^{2}+\zeta_{2}j_{\mu}j^{\mu})\\
\nonumber
&&-\partial_{\mu}\rho_{s}\partial^{\mu}\rho_{s}(\kappa_{d}+\alpha_{1}\rho_{s})
-\partial_{\mu}j_{\nu}\partial^{\mu}j^{\nu}(\zeta_{d}+\alpha_{2}\rho_{s})
-f_{v}(\partial^{\mu}j^{\nu})s_{\mu\nu}
.\end{eqnarray}
The parameters are given in Ref.~\cite{Rusnak:1997dj}, and arguments for
their naturalness in terms of effective field theory have also been presented
\cite{Furnstahl:2000rm}.

The relevant components of the symmetric energy momentum tensor in
the MFA are given by
\begin{eqnarray}
T^{++}_{MFA} & = & i\overline{\psi'}\gamma^{+}\partial^{+}\psi'\\
\nonumber T^{+-}_{MFA} & = &
  -2\overline{\psi'}(i\partial\!\!\!\!\!\:/-M)\psi'
 +2\rho_{s}^{2}(\kappa_{2}+\kappa_{3}\rho_{s}+\kappa_{4}\rho_{s}^{2})
\nonumber\\
 &&
+2j_{\mu}j^{\mu}(\zeta_{2}+\eta_{1}\rho_{s}+\eta_{2}\rho_{s}^{2}+\zeta_{4}j_{\nu}j^{\nu})
+\frac{i}{2}\overline{\psi'}(\gamma^{+}\partial^{-}+\gamma^{-}\partial^{+})\psi'
\end{eqnarray} and the Dirac equation is
\begin{eqnarray}
\label{diracfurn} \nonumber (i\partial\!\!\!\!\!\:/-M)\psi'
 & = & [\rho_{s}(2\kappa_{2}+3\kappa_{3}\rho_{s}+4\kappa_{4}\rho_{s}^{2})\\
&& \nonumber +\gamma_{\mu}j^{\mu}(2\zeta_{2}+2\eta_{1}\rho_{s}+2\eta_{2}\rho_{s}^{2}+4\zeta_{4}j_{\nu}j^{\nu})\\
&& +j_{\mu}j^{\mu}(\eta_{1}+2\eta_{2}\rho_{s})]\psi'
.\end{eqnarray}
Note that the various densities are constants within the
 MFA, we may define 
\begin{eqnarray} 
M^{*} & \equiv &
M+\rho_{s}(2\kappa_{2}+3\kappa_{3}\rho_{s}+4\kappa_{4}\rho_{s}^{2})
+j_{\mu}j^{\mu}(\eta_{1}+2\eta_{2}\rho_{s})\\
 G_{RF} & = &
\zeta_{2}+\eta_{1}\rho_{s}+\eta_{2}\rho_{s}^{2}+2\zeta_{4}j_{\mu}j^{\mu},
\end{eqnarray}
and follow Section \ref{walecka} to  obtain the equation 
\begin{equation}
\label{dirsolfurn}
(i\partial^{-}-2G_{RF}\widetilde{j}^{-})\psi_{+}
=\frac{k_{\bot}^{2}+(M^{*})^{2}}{k^{+}} \psi_{+},
\end{equation}
and the momenta
\begin{eqnarray}
\label{p-furn} \nonumber \frac{P}{\Omega}^{-} & = &
-2\kappa_{2}\rho_{s}^{2}-4\kappa_{3}\rho_{s}^{3}-6\kappa_{4}\rho_{s}^{4}
-2\eta_{1}\rho_{s}\rho_{B}^{2}-4\eta_{2}\rho_{s}^{2}\rho_{B}^{2}-4\zeta_{4}\rho_{B}^{4}\\
&&+\frac{4}{(2\pi)^{3}}\int_{F} d^{2}k_{\bot}dk^{+}\frac{k_{\bot}^{2}+(M^{*})^{2}}{k^{+}}\\
\label{p+furn} \frac{P}{\Omega}^{+} & = & 2G_{RF}\rho_{B}^{2} +
\frac{4}{(2\pi)^{3}}\int_{F} d^{2}k_{\bot}dk^{+}k^{+}
\end{eqnarray}
The derivation of  Eq. (\ref{p-furn}), involves adding and subtracting $M-M^{*}$
in order to use the Dirac operator
(\ref{dirsolfurn}) in obtaining the last term.
The numerical value of the effective mass is computed to be
$M^{*}=0.73M$ for $k_{F}=1.31 \text{ fm}^{-1}$.

The proof of the Hugenholtz-van Hove theorem is obtained by going
through the steps analogous to those of
Eqs.~(\ref{ea})-(\ref{hvhnum})             for this updated model.
One can see numerically in Table~~\ref{tab:summary} that the model
obeys the theorem. Additionally, the total energy, given by
substituting Eqs. (\ref{p-furn}) and (\ref{p+furn}) into Eq.
(\ref{ea}), is due entirely to nucleons, and therefore the model
must obey the Hugenholtz-van Hove theorem.


\begin{references}
\bibitem{simple} The intent of this paragraph is to capture the essence of the
  conventional nuclear binding effect without going into details. For example,
  scaling violations and related effects are neglected for  the
purpose of simplicity.
\bibitem{explain}
We neglect scaling violations and use the Bjorken limit.
  In Ref.\cite{jm} the quantity
$\chi(p,P)$
 is given as the Fourier transform of the a time-ordered
 product of nucleon fields. However, the more general
definition is in terms of  a covariant Green's function. Moreover,
the expression for the Green's function in terms of nucleon
 fields is not the same as the usual one if
light front dynamics are used\cite{yan12,yan34}.
Note also that $\chi(p,P)$ is $-i$ times a normalization factor times
 the Green's function.
\bibitem{nada} The derivation of the Hugenholtz-van Hove theorem
    in the context of the Walecka model must have been obtained before
                                our work. The relevance here is the
                                connection with the condition
                                $P^+=P^-$, and the connection with
                                the mesonic content of the nucleus.
\end{references}
\end{document}